\documentclass[aps, prl, twocolumn, superscriptaddress, showpacs]{revtex4}
\usepackage{graphicx}
\usepackage{dcolumn}
\usepackage{bm}
\usepackage{natbib}
\usepackage{upgreek}
\usepackage{amsmath}
\usepackage{amssymb}

\newcommand*{\vect}[1]{\mathbf{#1}}
\newcommand*{\cm}[1]{#1~cm$^{-1}$}
\newcommand*{\eIIa}{$e \| a$}
\newcommand*{\hIIb}{$h \| b$}
\newcommand*{\Dy}{DyMnO$_3$}

\begin{document}

\title{Electric field control of terahertz polarization in a multiferroic manganite with electromagnons}

\author{A. Shuvaev}
\author{V. Dziom}
\author{Anna Pimenov}
\author{M. Schiebl}
\affiliation{Institute of Solid State Physics, Vienna University of
Technology, A-1040 Vienna, Austria}
\author{A. A. Mukhin}
\affiliation{Prokhorov General Physics Institute, Russian Academy of
Sciences, 119991 Moscow, Russia}
\author{A. Komarek}
\author{T. Finger}
\author{M. Braden}
\affiliation{II. Physikalisches Institut, Universit\"{a}t zu
K\"{o}ln, 50937 K\"{o}ln, Germany}
\author{A. Pimenov}
\affiliation{Institute of Solid State Physics, Vienna University of
Technology, A-1040 Vienna, Austria}

\begin{abstract}

All-electrical control of a dynamic magnetoelectric effect is
demonstrated in a classical multiferroic manganite DyMnO$_3$, a
material containing coupled antiferromagnetic and ferroelectric
orders. Due to intrinsic magnetoelectric coupling with
electromagnons a linearly polarized terahertz light rotates upon
passing through the sample. The amplitude and the direction of the
polarization rotation are defined by the orientation of
ferroelectric domains and can be switched by static voltage. These
experiments allow the terahertz polarization to be tuned using the
dynamic magnetoelectric effect.

\end{abstract}

\date{\today}


\maketitle

Electric and magnetic field control of the propagation and the
polarization state of terahertz radiation is one of the
prerequisites for continuous progress of modern electronics. A
number of recent developments in this direction have been achieved
using multiferroics, i.e. materials simultaneously revealing
electric and
magnetic ordering~\cite{fiebig_jpd_2005,ramesh_nmat_2007,%
eerenstein_nature_2006,tokura_science_2006,cheong_nmat_2007}.
Several multiferroics provide not only a direct coupling between
static electric and magnetic properties but also give a possibility
to modify dynamic susceptibilities by external fields. Application
of a static magnetic field to the multiferroic materials leads to
dichroism in the terahertz range
\cite{pimenov_nphys_2006,kida_prb_2011} or even to more complex
effects like controlled chirality \cite{bordacs_nphys_2012} or
directional dichroism
\cite{kezsmarki_prl_2011,takahashi_nphys_2012,takahashi_prl_2013}.
Electric control of terahertz radiation is more difficult to realize
and it has been recently demonstrated in Raman scattering
experiments \cite{rovillain_nmat_2010}.

Dynamical properties of several multiferroic materials in the
terahertz range are governed by novel
magnetoelectric modes called electromagnons~\cite{tokura_phil_2011,shuvaev_jpcm_2011,%
sushkov_jpcm_2008,smolenskii_ufn_1982}. Electromagnons may be
defined as collective excitations of the magnetic structure which
are coupled to the electric dipole moment.They may be regarded as a
mixture of magnons and phonons.
In orthorhombic rare earth manganites RMnO$_3$ one generally observes several electromagnons in the terahertz and sub-terahertz range. A strong high frequency mode around 2-3~THz is well understood on the basis of a symmetric Heisenberg exchange (HE) coupling
\cite{aguilar_prl_2009,lee_prb_2009} as a zone edge magnon which can be excited by electric component of the electromagnetic radiation. A second intensive mode existing at 0.5-1~THz has been explained using the same mechanism but including a Brillouin zone folding due to modulation of the magnetic cycloid~\cite{lee_prb_2009,stenberg_prb_2009}. In the sub- terahertz frequency range a series of weaker modes is observed in optical~\cite{pimenov_jpcm_2008,shuvaev_jpcm_2011} and neutron scattering experiments~\cite{senff_jpcm_2008}. These modes are explained as the magnetic eigenmodes of the spin cycloid in RMnO$_3$. Some of these modes may get an electrical dipole activity due to the relativistic Dzyaloshinskii-Moriya (DM) mechanism. Dynamic contributions due to this mechanism
have been investigated both experimentally and
theoretically~\cite{katsura_prl_2007,pimenov_jpcm_2008,pimenov_prl_2009,cano_prb_2009,senff_prl_2007}.
In spite of its weakness, the DM interaction is a promising
mechanism especially in application to spiral magnets as it connects
static spontaneous polarization and magnetic
structure~\cite{katsura_prl_2007,mostovoy_prl_2006}. This mechanism
is responsible for the switching of ferroelectric polarization by
magnetic field and for the control of magnetic structure by electric
voltage in spiral magnets~\cite{cheong_nmat_2007}. It may be
expected that in the frequency range where the dynamics is governed
by the DM mechanism, the terahertz light will be controlled by
electric field as well. In present experiments we utilize this idea
for two purposes: we obtain a direct evidence of dynamical
magnetoelectric coupling within the DM electromagnon and we
demonstrate a possibility to control the polarization of terahertz
light by applying static electric fields.

\Dy{} is a multiferroic manganite with orthorhombic structure. The
high-temperature paramagnetic state in this material transfers into
an incommensurate antiferromagnetic structure below $T_\textrm{N}
\simeq 39$~K. At lower temperatures a second phase transition into a
ferroelectric phase takes place at $T_c \simeq 19$~K. By analogy to
TbMnO$_3$ this phase is most probably a cycloidal
antiferromagnet~\cite{kenzelmann_prl_2005} with an incommensurate
propagation vector. Below the transition to the cycloidal state
\Dy{} reveals static electric polarization which is aligned along
the $c$-axis ($Pbnm$ crystallographic setting is used throughout
this paper). This polarization is well described by the DM coupling
which leads to a simple expression~\cite{mostovoy_prl_2006}:
\begin{equation}
\vect{P}_0 \sim  \vect{\delta}_{j \rightarrow j + 1} \times (
\vect{S}_j \times \vect{S}_{j+1}) \quad. \label{p}
\end{equation}

\begin{figure}
\begin{center}
\includegraphics[width=0.95\linewidth, clip]{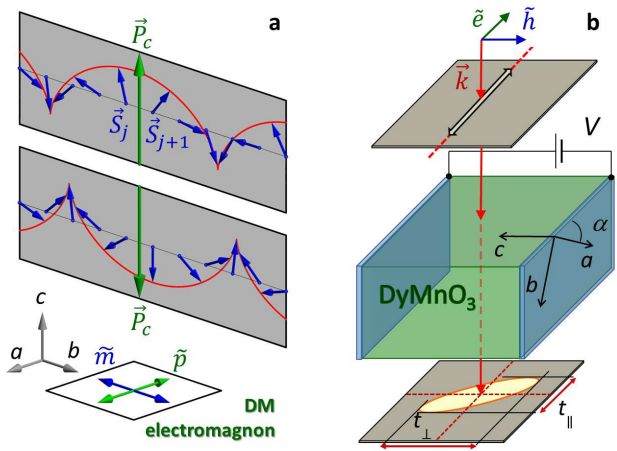}
\end{center}
\caption{\emph{Experiment to observe electrically controlled dynamic
magnetoelectric effect.} \textbf{a} - Schematic representation of
the magnetic $bc$-cycloid and the static electric polarization
(green arrows) in \Dy{}. Shown are two possible domains with
opposite orientations of the cycloid and the polarization. Bottom
diagram indicates oscillations of electric and magnetic moments for
magneto-electrically active mode (DM electromagnon). \textbf{b} -
Geometry of the \Dy{} crystal and of the experimental apparatus to
separate waves of different polarizations.} \label{fig1}
\end{figure}

Here $\vect{S}_j$ and $\vect{S}_{j+1}$ are the neighbor Mn$^{3+}$
spins within $ab$-planes and $\vect{\delta}_{j \rightarrow j + 1}$
is the vector connecting them (see Fig.~\ref{fig1}\textbf{a}). The
spin cycloid breaks the space inversion symmetry and has two
possible rotation directions of the spins ($\circlearrowleft$ and
$\circlearrowright$). According to Eq.~(\ref{p}), the sign of the
static polarization is opposite in these two cases (see Fig.
1\textbf{a}). Therefore, the antiferromagnetic domains are
simultaneously ferroelectric domains, and the orientation of the
spin cycloid is also affected by external electric field.

The idea of the present experiment is based on the DM coupling
between static and dynamic properties in \Dy{}. A schematic picture
of the cycloidal magnetic structure in \Dy{} is shown in
Fig.~\ref{fig1}\textbf{a}. Due to an incommensurate character of the
cycloid, the solution of the dynamic equations for this structure
reveals three eigenmodes (see Supplementary Information for more
details). For the present experiment only one mode is the most
promising. Within this mode magnetization and electric polarization
oscillate along the $b$ and $a$ axes, respectively (DM electromagnon
in Fig.~\ref{fig1}\textbf{a}). Therefore, this mode can be excited
either via electric channel by \eIIa{} and via magnetic channel by
\hIIb{}. Moreover, these two channels are not independent. The
electric excitation drives also the magnetic moment and vice versa.
As discussed in more details in the Supplementary Information, this
cross coupling is manifested in the existence of the nonzero dynamic
magnetoelectric susceptibility $\chi_{ab}^{me}$.

The main experimental difficulty to observe dynamic magnetoelectric
effect in \Dy{} is that it cannot be detected in an experiment with
an $ab$-plane cut crystal. In such geometry the ac fields of the
incident wave are either $e \| b$ and $h \| a$ and do not excite the
electromagnon at all, or they are \eIIa{} and \hIIb{} and,
therefore, they both excite the electromagnon at the same time. The
existence of the magnetoelectric effect in such geometry does not
lead to an emergence of a wave with the perpendicular polarization
but only slightly changes the absorption of light. In order to
overcome this difficulty, the sample with tilted axes has to be
used. The geometry of such an experiment is shown in
Fig.~\ref{fig1}\textbf{b}. In the following arguments we assume
incident wave with electric field component $e \| ab$-plane of the
crystal which excite the DM electromagnon via electric channel. This
geometry is equivalent to $e \bot c$ in Fig.~\ref{fig1}\textbf{b}
and contains both components of the electric field  $e \| a$ and $e
\| b$. Because the DM electromagnon has nonzero magnetoelectric
component $\chi_{ab}^{me}$, an ac magnetic field \hIIb{} will be
induced by this excitation. This electromagnetic field corresponds
to a wave with polarization perpendicular to the incident wave with
$h \| c$. Thus, an appearance of a signal in crossed polarizers is a
characteristic of a nonzero magnetoelectric susceptibility. These
qualitative arguments are supported by rigorous calculations given
in the Supplementary Information.

We note that within the present experiment the existence of DM
electromagnon which can be excited at the center of the Brillouin
zone is crucial. As shown in the rigorous solution Supplementary
Eq.~(5), the electrically and magnetoelectrically active mode can be
represented as a symmetric superposition of two magnons with
wavevectors $\vect{q} = +\vect{Q}$ and $\vect{q} = -\vect{Q}$. Here
$\vect{Q}$ is the modulation vector of the magnetic cycloid
\cite{aguilar_prl_2009}. The symmetric mode represents an
electromagnon which have nonzero dynamic polarization along the $x$
axis and, therefore, can be excited by electromagnetic wave with $e
\| a$.

In case of (although much stronger) Heisenberg
electromagnons~\cite{aguilar_prl_2009,lee_prb_2009} which are
excited as a zone edge magnons, the present experiment would not
work. For the zone edge electromagnon the neighbor spins oscillate
out-of-phase, which cancels the resulting magnetic moment. Although
this mode reveals strong electric contribution, the magnetic and
magnetoelectric susceptibilities are zero. As will be shown in more
detail below (Fig.~\ref{fig3}), the dynamic magnetoelectric effects
observed in \Dy{} are indeed centered around the weak DM
electromagnon at 210~GHz and they are absent around strong
Heisenberg electromagnon around 550~GHz.

The remaining point is the requirement of an electrical poling of
\Dy{} crystal. Without poling, two types of domains
(Fig.~\ref{fig1}\textbf{a}) coexist in the sample. The domains with
the opposite ($\circlearrowleft$ or $\circlearrowright$) rotation of
the spin cycloid reveal the opposite sign of the magnetoelectric
susceptibility, canceling the effect. In order to avoid the signal
compensation from different domains, the sample has been poled in
static electric fields $E||c$. Such poling orients the majority of
the domains along one direction.

\begin{figure}
\begin{center}
\includegraphics[width=0.85\linewidth, clip]{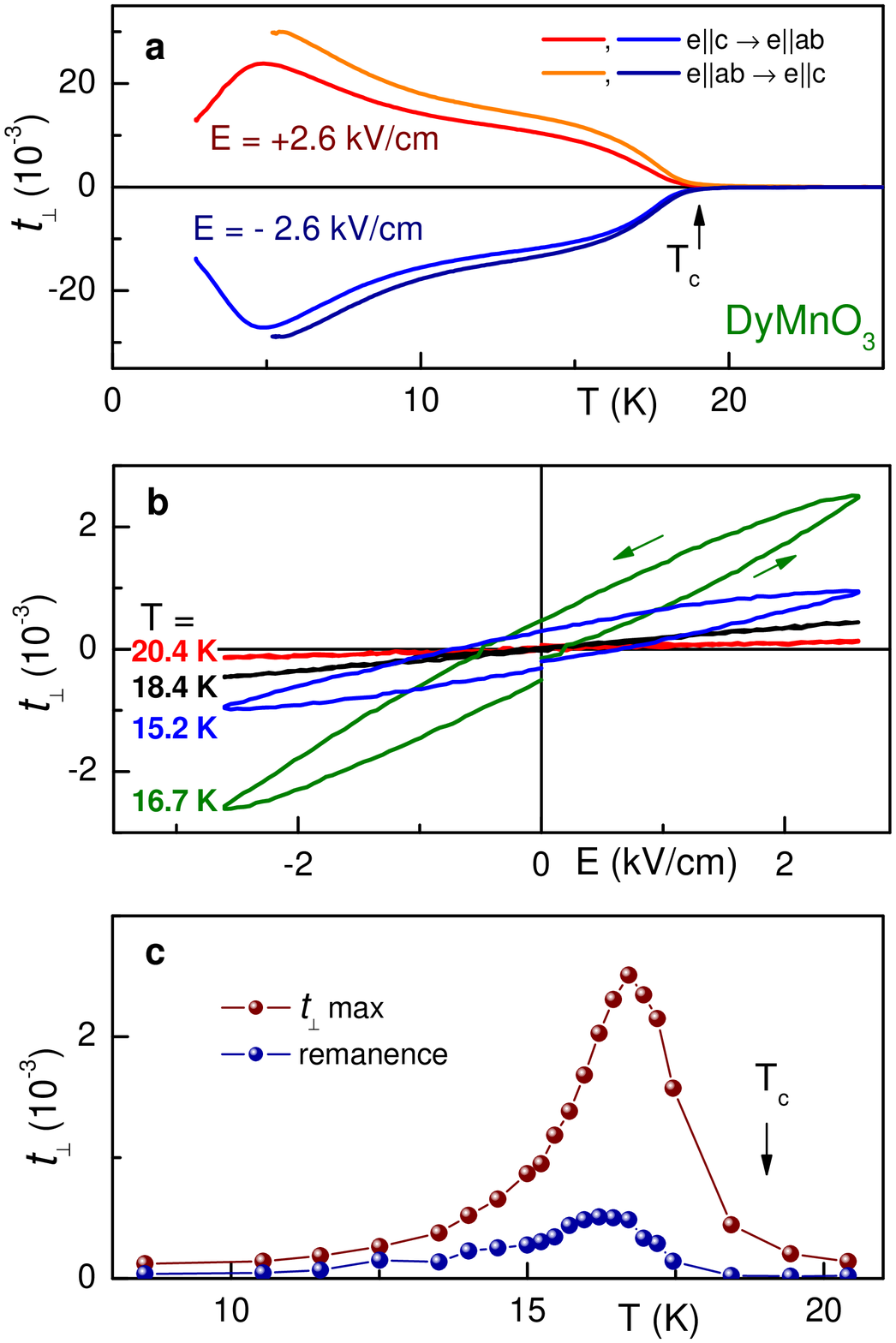}
\end{center}
\caption{\emph{Controlling of terahertz light by static electric
field in \Dy{}.} \textbf{a} - Transmitted  terahertz signal in \Dy{}
at $\nu=210$~GHz in crossed polarizers for different polarizations
and poling electric fields (field-cooling). The geometry of the
experiment is given in Fig.\ref{fig1}\textbf{b}. The notation $e \|
ab$ is equivalent to $e \bot c$ in Fig.\ref{fig1}\textbf{b}.
Positive and negative sign of $t_{\bot}$ reflects clockwise and
counterclockwise polarization rotation, respectively. Arrow
indicates the phase transition to the ferroelectric phase.
\textbf{b} - Electric voltage dependencies of the transmission in
crossed polarizers for various temperatures and for the zero-field
cooled (ZFC) sample. \textbf{c} - Maximum available signal in
crossed polarizers and the remanence signal as function of
temperature in ZFC case. Symbols -- experiment, lines are to guide
the eye.} \label{fig2}
\end{figure}

Figure~\ref{fig2}\textbf{a} shows a typical result of the experiment
in crossed polarizers geometry. We note that crossed polarizers
separate the incident polarization from the induced one. As
expected, no signal could be be detected in the paraelectric phase.
Immediately upon the onset of the ferroelectric phase, distinct
polarization rotation is observed with the sign of the signal
correlating with the sign of the static field
(Fig.~\ref{fig2}\textbf{a}). Here we plot clockwise rotation of the
polarization as positive signal and the counterclockwise rotation as
negative signal. Equivalently, the positive and negative sign of
$t_{\bot}$ reflect the 180$^\circ$ phase difference between the
experimental signal for different sign of static electric field. No
signal is observed without poling of the sample. These results
demonstrate the validity of the qualitative arguments given above.

Another important result of this work is shown in
Figs.~\ref{fig2}\textbf{b},\textbf{c}. Here, not far from the phase
transition into the ordered state, the ferroelectric domains may be
switched by moderate static field. Due to direct coupling of static
and dynamic properties, the sign of the magnetoelectric
susceptibility is switched as well. Therefore, in this range we can
directly influence the signal $t_{\perp}$ and the polarization
rotation of the terahertz radiation by static electric field.

A significant difference between the experiments in
Figs.~\ref{fig2}\textbf{a},\textbf{c} is that a field-cooling
experiment is performed in the fist case and a zero-field-cooled
experiment in the second case. Because in the field-cooled case the
sample is cooled starting from the paraelectric state, it is much
easier to align the ferroelectric domains by static field. In the
zero-field-cooled sample the electric domains are not oriented.
Especially at low temperatures the coercive field is strong and the
static electric field cannot reorient the domains. The reorientation
of the domains takes place close to the ferroelectric transition
only, which explains the maxima observed in
Figs.~\ref{fig2}\textbf{c}. Finally, we note that the effects in
Figs.~\ref{fig2}\textbf{a},\textbf{b} are due to the same
microscopic mechanism, but a direct switching of polarization in
Fig.~\ref{fig2}\textbf{b} is more relevant from the point of view of
possible applications.

\begin{figure}
\begin{center}
\includegraphics[width=0.85\linewidth, clip]{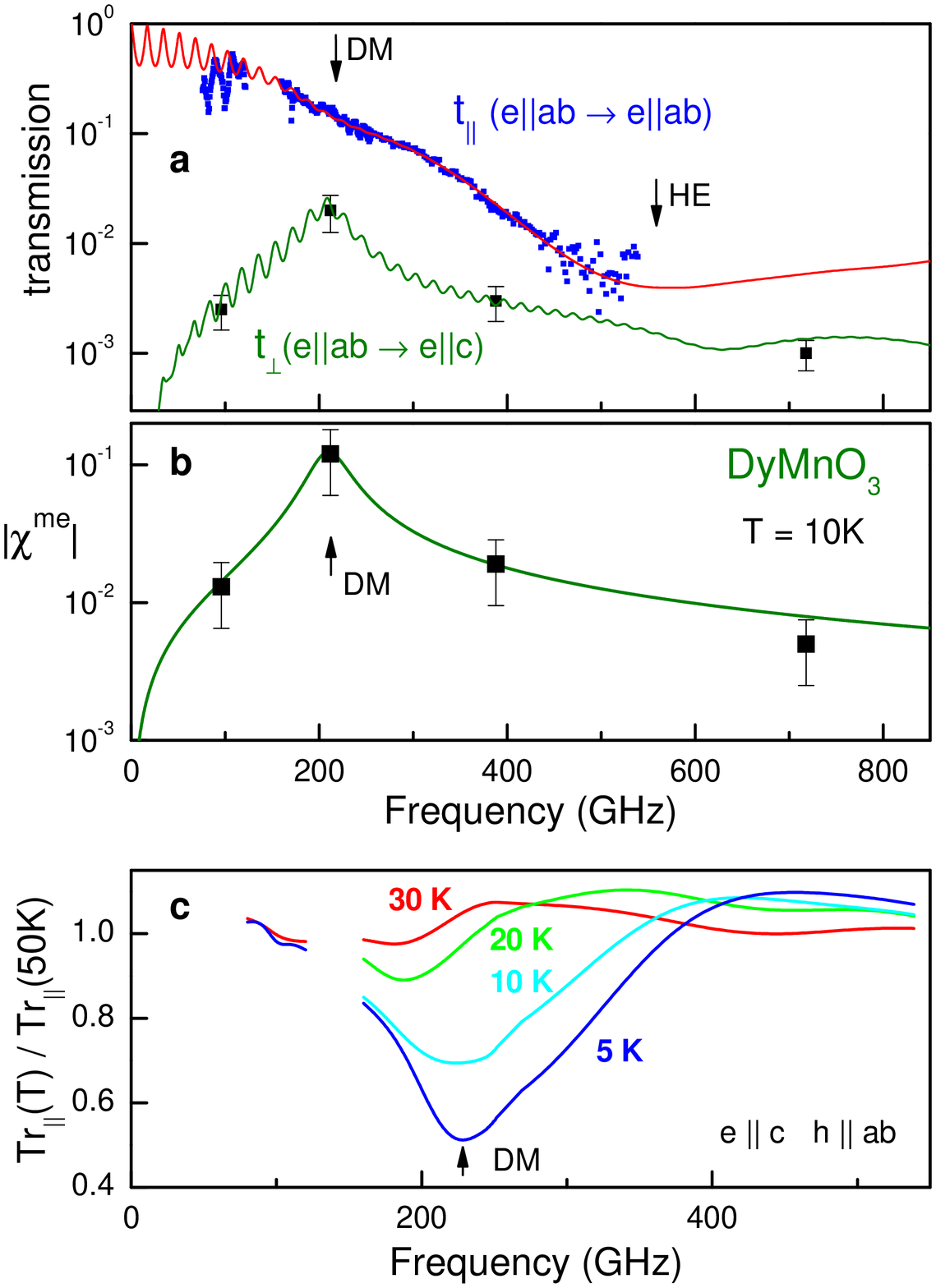}
\end{center}
\caption{\emph{Electromagnons in \Dy{}.} \textbf{a} - Transmission
spectra of \Dy{} in parallel (blue symbols) and crossed (black
symbols) polarizers. The transmission is dominated by the Heisenberg
electromagnon at 550~GHz (marked as \textrm{HE}). Much weaker
Dzyaloshinskii-Moriya electromagnon (marked as \textrm{DM}) at
210~GHz is responsible for the observed dynamic magnetoelectric
effect and for nonzero signal in crossed polarizers. Symbols -
experiment, lines are fits according to Fresnel optical equations.
\textbf{b} - Magnetoelectric susceptibility as obtained from the
spectra in \textbf{a}. \textbf{c} - Transmission in parallel
polarizers and in the transparent geometry with $e\|c$ showing a
magnetically excited DM electromagnon.} \label{fig3}
\end{figure}

In order to prove the proposed mechanism of the polarization
rotation, a series of spectroscopic experiments has been carried
out. The terahertz dynamics in our frequency range is dominated by a
strong electromagnon at about 550~GHz (\cm{18}). This electromagnon
is responsible for a relatively low transmission in the geometry
with \eIIa{}, seen as blue symbols in Fig. 3\textbf{a}. This
excitation most probably originates from a symmetric Heisenberg
exchange
mechanism~\cite{aguilar_prl_2009,lee_prb_2009,kida_prb_2008} and it
does not contribute to the effects described in this work.

In the transmission spectra \eIIa{} another weaker excitation can be
seen close to 210~GHz. This mode is observed both in the geometry
\eIIa{} (Fig.~\ref{fig3}\textbf{a}, red curve) as well as in the
perpendicular geometry $e\|c$ (Fig.~\ref{fig3}\textbf{c}). In the
latter geometry the sample is more transparent as the main
absorption mechanism due to the Heisenberg exchange with the
component \eIIa{} is absent. In close analogy to a similar spectral
analysis~\cite{pimenov_prl_2009} in TbMnO$_3$, we attribute the
210~GHz mode to the zone-center eigenmode of the cycloidal
structure. This mode gets its intensity predominantly due to the
Dzyaloshinskii-Moriya mechanism. Because static electric
polarization is governed by the same mechanism, static and dynamic
properties are strongly correlated for the 210~GHz mode. As
discussed above, this connection is the basic mechanism to produce
electrically controlled rotation of the terahertz polarization.

As presented in more details in the Supplementary Materials, the
210~GHz mode of the cycloidal spin structure reveals nonzero
electric $\chi_a^e$, magnetic $\chi_b^m$, and magnetoelectric
$\chi_{ab}^{me}$ susceptibilities. This mode can be excited by both,
an ac electric field \eIIa{} and ac magnetic field \hIIb{} and can
be therefore called a DM electromagnon. In agreement with these
arguments, a rotation of the polarization is the strongest close to
210~GHz and fades away on both sides of the resonance. This result
is shown in Fig.~\ref{fig3}\textbf{a} with black squares. Green
solid line represents the result of calculations of the transmission
in crossed polarizers assuming Lorentz line shape of the DM
electromagnon at 210~GHz. Tiny oscillations in this curve reflect
the Fabry-P\'{e}rot resonances on the sample surfaces. In order to
obtain the magnetoelectric susceptibility directly from the measured
transmission, the complex transmission matrix has been inverted
numerically.  The frequency dependence of a resulting
magnetoelectric susceptibility in \Dy{} is shown in
Fig.~\ref{fig3}\textbf{b} by black symbols. We note that in spite of
the complexity of the data treatment, a nonzero signal in crossed
polarizers is to a leading term directly proportional to
$\chi^{me}$~\cite{shuvaev_epjb_2011}. This explains qualitative
similarity of the frequency dependencies of $t_\bot(\nu)$ and
$\chi^{me}(\nu)$ in Fig.~\ref{fig3}\textbf{a},\textbf{b}.

From the Lorentzian fits in Fig. 3 the intensities of the DM
electromagnon are obtained as follows: electric contribution (Fig.
3\textbf{a}) $\Delta \varepsilon_a = 1.7 \pm 0.3$; magnetic
contribution (Fig. 3\textbf{c}) $\Delta \mu_b = 0.010 \pm 0.002$;
Magnetoelectric contribution: (Fig. 3\textbf{b}) $\Delta
\chi_{ab}^{me} = 0.03 \pm 0.01$. We see that the universality
condition Supplementary Eq. (10) is not fulfilled in \Dy:
$\sqrt{\Delta \varepsilon_a \Delta \mu_b}=0.13 > \chi_{ab}^{me}$.
This disagreement most probably indicates that large part of the DM
electromagnon spectral weight is provided by the Heisenberg exchange
mechanism. Indeed, a theoretical estimate of electric contribution
\cite{aguilar_prl_2009} from Supplementary Eq.~(8) gives the value
$\Delta \varepsilon_a \sim 0.2$ substantially smaller than the
experimental result.

In orthorhombic rare earth manganites (RMnO$_3$, R = Dy,
Tb, Eu:Y) strong zone edge electromagnons in the terahertz spectra
are due to symmetric Heisenberg exchange mechanism. However, their
properties do not correlate with the behavior of static electric
polarization, because the latter is due to antisymmetric
Dzyaloshinskii-Moriya coupling. On the contrary, in present
experiments static and dynamic properties are controlled by the same
DM mechanism, which explains the observed voltage control of
terahertz light.

Finally, the observed results differ from such well-known effect like electro-optical modulation\cite{landau_book8} (Pockels effect). Several arguments support this statement: i) the observed signal qualitatively follow the ferroelectric polarization (Fig.~\ref{fig2}\textbf{a}) and disappear in unpoled sample at low temperatures (Fig.~\ref{fig2}\textbf{c}); ii) in poling experiment the same signal is observed if only half as intensive electric voltage is applied, i.e. the effect saturates in field; iii) the frequency dependence of the observed magnetoelectric signal follows the Lorentzian line shape of the DM electromagnon.

In conclusion, we investigate dynamic magnetoelectric effect based
on DM electromagnon in \Dy. Due to off-diagonal elements of the
magnetoelectric susceptibility a polarization plane rotation of the
transmitted radiation is observed. The amplitude and the direction
of the polarization rotation can be controlled and switched by
static electric voltage. From the spectral analysis a full set of
magnetic, electric, and magnetoelectric susceptibilities of the DM
electromagnon in \Dy is obtained.

\subsection*{Acknowledgements}
We thank K. Hradil for the help in sample orientation. This work was
supported by the by the German Research Foundation DFG, by Russian
Foundation for Basic Researches (N 12-02-01261), and by the Austrian
Science Funds (I815-N16, W1243).

\bibliographystyle{unsrt}
\bibliography{literature}

\end{document}